\documentclass[a4paper]{article}

\usepackage{ISCSLP2024}

\title{RobustSVC: HuBERT-based Melody Extractor and Adversarial Learning for Robust Singing Voice Conversion}
\name{
	{Wei Chen, Xintao Zhao, Jun Chen, \\
 Binzhu Sha, Zhiwei Lin, Zhiyong Wu$^{*}$\thanks{$^{*}$ Corresponding author.}}
}
\address{
  {
  	Shenzhen International Graduate School, Tsinghua University, Shenzhen, China
  }
}

\email{
	{
        chenw23$@$mails.tsinghua.edu.cn, zywu$@$sz.tsinghua.edu.cn
        }
}
\usepackage{multirow}
\usepackage{makecell}
\usepackage{amsmath}

\begin{document}

\maketitle
\begin{abstract}
  Singing voice conversion (SVC) is hindered by noise sensitivity due to the use of non-robust methods for extracting pitch and energy during the inference. 
As clean signals are key for the source audio in SVC, music source separation preprocessing offers a viable solution for handling noisy audio, like singing with background music (BGM). 
However, current separating methods struggle to fully remove noise or excessively suppress signal components, affecting the naturalness and similarity of the processed audio. 
To tackle this, our study introduces RobustSVC, a novel any-to-one SVC framework that converts noisy vocals into clean vocals sung by the target singer. 
We replace the non-robust feature with a HuBERT-based melody extractor and use adversarial training mechanisms with three discriminators to reduce information leakage in self-supervised representations. 
Experimental results show that RobustSVC is noise-robust and achieves higher similarity and naturalness than baseline methods in both noisy and clean vocal conditions.
\end{abstract}
\noindent\textbf{Index Terms}: singing voice conversion, adversarial training, noise-robust

\vspace{-8pt}
\section{Introduction}

SVC, a downstream research task of voice conversion (VC), aims to transform the timbre of the original singer's voice in a song to that of a target singer without altering aspects like melody and lyrics. 
Compared to conventional VC task, the SVC system requires a more sophisticated modeling of acoustic features. 
In the VC task, slight differences in prosody and tempo between the converted audio and input audio are acceptable. 
However, within the realm of SVC, musical features such as melody are closely tied to the song's essence, which means they are song-dependent and should be precisely preserved.

Traditional SVC methods ~\cite{kobayashi2014statistical,villavicencio2010applying,kobayashi2015statistical,toda2007one} generally use statistical models to build the mapping between the parallel samples which are expensive in practice. 
Non-parallel SVC is a more challenging but more practical undertaking in applications. 
A common way to achieve non-parallel SVC is extracting a singer-independent content representation and then producing the converted singing voices by utilizing the target singer embedding. 
To get content representations, some methods tried to disentangle content feature from speech in an unsupervised manner~\cite{luo2020singing,deng2020pitchnet,lu2020vaw}. 
Nevertheless, due to the difficulty of unsupervised learning, its ability of feature disentanglement is not robust enough for redundant noise from input data~\cite{li2021ppg}.


\begin{figure}[t]
  \centering
  \includegraphics[width=\linewidth]{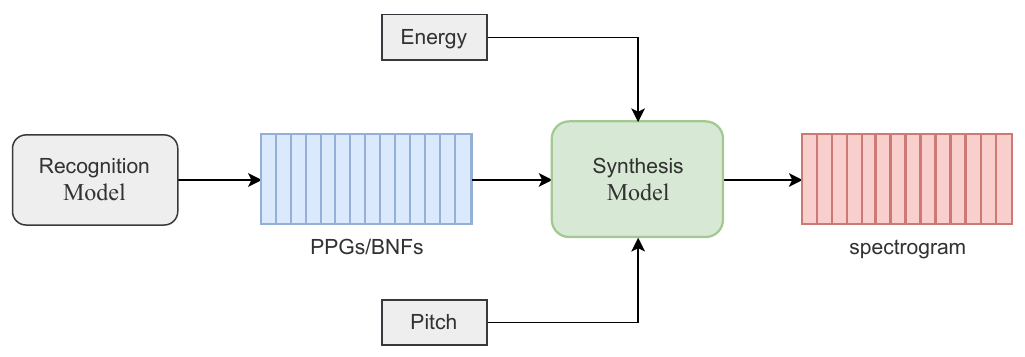}
  \caption{Recognition-synthesis based SVC framework}
  \label{fig:baseline}
  \vspace{-15pt}
\end{figure}

Recognition-synthesis based SVC is another representative approach of the non-parallel SVC, demonstrating significant performance enhancement~\cite{li2021ppg,chen2019singing,guo2020phonetic,zhang2023leveraging,ning2023vits,wang2021towards}.
The framework is shown in Figure~\ref{fig:baseline}.
An intermediate embedding is extracted from a certain layer of the ASR model, either phonetic posteriorgrams (PPGs)~\cite{chen2019singing} from the last layer regardless of preformed extractor structure, or bottleneck features (BNFs)~\cite{ning2023vits} from the penultimate layer with well-designed dimension, and then fed into a SVC system after combining with pitch and energy.

Since the singing voice contains much more characteristics like melody, which are difficult to model, several works have tried to extract melody feature while all of these efforts require clean source audio signals.
In ~\cite{li2021ppg}, in addition to pitch, Mel-spectrogram is also inputted into the reference encoder to implicitly capture singing characteristics. 
Nonetheless, Mel-spectrogram contains a considerable amount of singer-specific information that is challenging to eliminate, ultimately leading to performance degradation. 
Moreover, in the process of dealing with noisy audio, such as singing alongside BGM, extracting accurate melody information from Mel-spectrogram presents significant challenges.
In ~\cite{wang2021towards}, they utilized the HuBERT feature and pitch for melody feature modeling. 
However, as the pitch is not a robust feature, when confronted with noisy singing voice inputs, the effectiveness of melody extraction will significantly decrease, consequently affecting the quality of converted audio.
Despite using a music source separation model for preprocessing before conversion, it can only partially alleviate the noise sensitivity concern. 
Specifically, existing separating methods suffer from residual noise or over-suppression problems, leading to loss of voice information~\cite{dai2020noise} and subsequently affecting the similarity and naturalness of the converted audio.
\begin{figure*}[t]
  \centering
  \includegraphics[width=\linewidth]{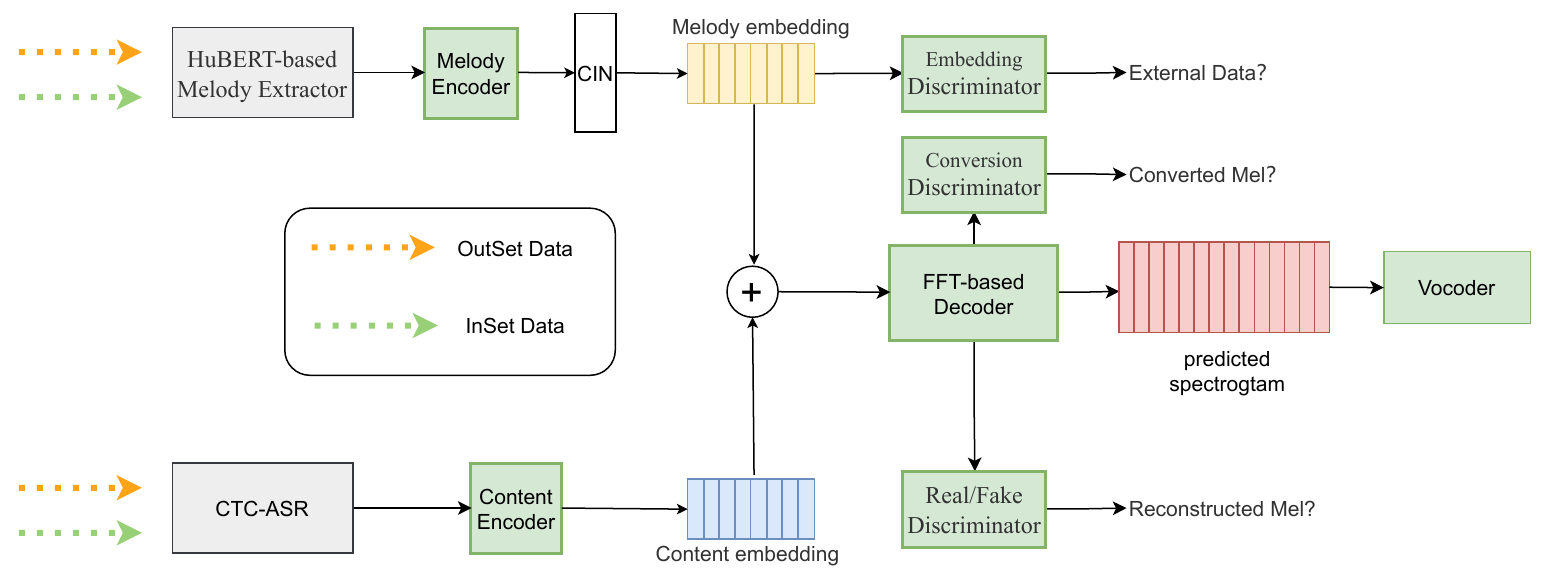}
  \caption{RobustSVC framework.  }
  \label{fig:arc}
  \vspace{-15pt}
\end{figure*}



In this work, to address this challenge, we propose RobustSVC, a high-quality noise-robust SVC system capable of converting noisy source audio into clean audio performed by the target singer.
Since the HuBERT feature contains melody information mentioned in ~\cite{wang2021towards}, we design a HuBERT-based melody extractor with the purpose of modeling melody to further address the noise sensitivity issue. 
In order to provide representations with good disentangling properties~\cite{zhao2022disentangling}, RobustSVC adopts the BNFs extracted from the ASR model trained with Connectionist Temporal Classification loss (CTC-BNFs). 
Additionally, three auxiliary discriminators are implemented to generate Mel-spectrogram with high similarity as well as quality.
The contributions of our work are as follows:

\begin{itemize}
\item We propose a HuBERT-based melody extractor that does not rely on non-robust methods for extracting pitch and energy during inference, achieving more robust and accurate melody modeling. 
\item We further develop a novel SVC framework, RobustSVC, which combines CTC-BNFs and an adversarial training strategy to reduce residual speaker information, enhance audio quality, and improve voice similarity.
\item Experimental results reveal that RobustSVC is not only noise-robust but also outperforms the baseline method in terms of subjective and objective evaluations for both noisy and clean vocal conditions, achieving higher similarity and naturalness.
\end{itemize}

\section{Method}
Our proposed RobustSVC system, shown in Figure~\ref{fig:arc}, consists of six components: (i) A content encoder to encode CTC-BNFs into content embedding. 
(ii) A HuBERT-based melody extractor to accurately model the melody. 
(iii) A melody encoder followed by conditional instance normalization (CIN)~\cite{dumoulin2016learned} to encode melody feature. 
(iv) A Feed Forward Transformer (FFT) based decoder. 
(v) three different discriminators which could help the framework generate Mel-spectrogram with high similarity as well as quality. 
(vi) A pre-trained HiFi-GAN~\cite{kong2020hifi} vocoder which is finetinued by Ground Truth Aligned (GTA) mode. 

The HuBERT-based melody extractor and ASR system are initially trained and then frozen to extract melody feature and CTC-BNFs, respectively. 
Afterwards, they are input into the encoder-decoder model to be trained in generating high-quality Mel-spectrogram, which is then utilized by the HiFi-GAN vocoder for fine-tuning to produce the converted audio. 
The details of each part are as follows.
\subsection{ASR System}
The ASR acoustic model based on Conformer~\cite{gulati2020conformer} achieves excellent recognition performance by combining the global modeling ability of the attention mechanism with the local modeling ability of CNN. 
We train the ASR model using the Conformer encoder with the CTC loss function and extract 256-dims BNFs from the last hidden layer of the Conformer encoder to utilize its effective disentangling properties, which lead to high similarity~\cite{zhao2022disentangling}.

\subsection{Encoder-Decoder Model}

As shown in Figure~\ref{fig:arc}, the proposed conversion model is an encoder-decoder architecture based on the FFT, renowned for its efficient computational speed and high-quality output generation.
First, the input melody feature is processed by the melody encoder with 3 FFT blocks. 
The output of the melody encoder is concatenated with the content embedding extracted by the content encoder after eliminating the timbre information through the CIN module. 
The concatenated content and melody embedding are further input into the decoder based on the 6 FFT blocks to generate high-quality spectrograms. 
We compute the reconstruction loss between the ground truth and the reconstructed Mel-spectrogram, which is defined as follows:
\begin{equation}
    L_{rec} =  \Vert y^{f} - y^{g} \Vert_1
\end{equation}
where $y^{g}$ and $y^{f}$ are the Mel-spectrogram of the ground truth and reconstructed audio respectively.

\subsection{HuBERT-based Melody Extractor}
\begin{figure}[ht]
  \centering
  \includegraphics[width=0.8\linewidth]{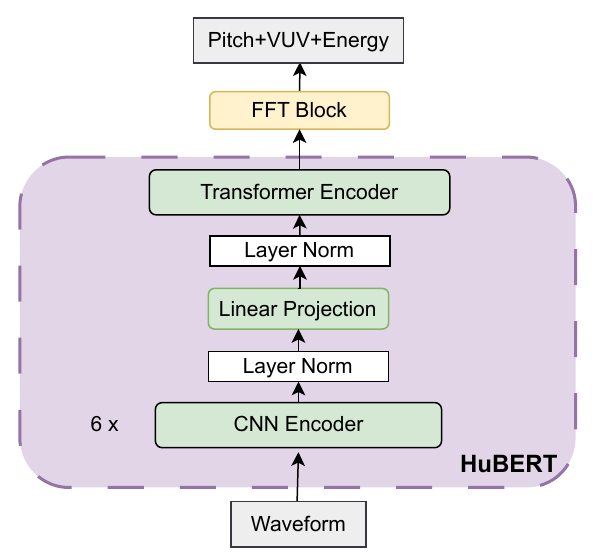}
  \caption{HuBERT-based melody extractor}
  \label{fig:hub}
  \vspace{-15pt}
\end{figure}

To avoid relying on non-robust methods for extracting pitch and energy in obtaining melody information, we develop a melody extractor based on the HuBERT architecture, as depicted in Figure~\ref{fig:hub}.
This extractor comprises a pre-trained HuBERT model augmented with three FFT blocks, which is fine-tuned for melody extraction tasks within limited vocal data scenarios. 
To strengthen the model's noise robustness, noisy training data is fed into the model. 
The output from the penultimate layer of the melody extractor is chosen as the melody feature input for the melody encoder. 
The loss function of the melody extractor is defined as follows:
\begin{equation}
    L_{melody} =  \Vert P^{f} - P^{g} \Vert_1 + \Vert E^{f} - E^{g} \Vert_1 +\Vert V^{f} - V^{g} \Vert_1
\end{equation}
where $ P^{f}$ and $P^{g}$ are the ground truth and predicted pitch, $ E^{f}$ and $E^{g}$ are the ground truth and predicted energy, $V^{f}$ and $V^{g}$ are the ground truth and predicted voice/unvoice flag (VUV) respectively.

\subsection{Discriminator Group}
Inspired by~\cite{10219976}, we introduce three different discriminators.
The real/fake discriminator aims to determine whether a sequence of Mel-spectrogram is extracted from the ground-truth audio or reconstructed by the SVC model, enhancing the quality of Mel-spectrogram generation.
Notably, the source audio used for reconstructed Mel comes from the target speaker’s corpora. 
The conversion discriminator is to distinguish whether a sequence of Mel-spectrogram has been converted by the SVC model, whose source audio comes from the external corpora. 
The embedding discriminator is tasked with judging whether a sequence of melody embeddings is extracted from the external corpora, which contain speaker information different from the target speaker. 
The latter two discriminators are able to minimize residual speaker information across different domains.
The final loss functions are given as follows:
\vspace{-1.2pt}
\begin{align}
    L(D)&=L_{sim}(D)+L_{rf}(D_{r}) \\
    L(G)&=L_{sim}(G)+L_{rf}(G)+L_{rec}
\end{align}
\vspace{-1.2pt}
where $ L_{sim}$ is the similarity-adversarial loss of conversion discriminator and embedding discriminator, $L_{rf}$ is the loss of real/fake discriminator aimed at enhancing the quality of the generated spectrograms.

\subsection{Vocoder}
As the pre-trained HiFi-GAN~\cite{kong2020hifi} vocoder is trained on pure speech data, its reconstruction performance on singing waveforms is limited. 
To address this limitation, we integrate the HiFi-GAN vocoder model into the training pipeline after a well-trained SVC model, a process known as the GTA mode~\cite{shen2018natural}. 
During the finetuning process, the parameters of the SVC model are frozen, and the vocoder leverages the spectra predicted by the SVC as input to map them to real waveform data.
This strategy aims to mitigate any negative effects arising from differences in the training data domain while also smoothing out deficiencies in the predictions made by the SVC model.

\section{Experiment}

\subsection{Experiment Setup}
The SVC experiments are conducted on an Opencpop~\cite{wang2022opencpop} corpus performed by a professional female singer, totaling 5.2 hours in duration. 
The training data is augmented by varying the speaking rate from 0.9 to 1.5 times to enhance melody diversity. 
ASR system is trained with the same configuration as described in~\cite{zhao2022disentangling}, and 256-dimensional BNFs are extracted from the ASR model.
Additionally, we select a 20-hour subset of the OpenSinger~\cite{huang2021multi} corpus as our external unannotated dataset for training. 
During the training steps, the weight of $L_{sim}$ and $L_{rf}$ is set to 0 until reaching 50,000 steps.

In the training of the melody extractor, we utilize an almost 30-hour subset of OpenSinger and MUSDB~\cite{rafii2017musdb18} for finetuning, resulting in a combined total duration of around 40 hours. 
The rest of the OpenSinger corpus is served as test data. 
Throughout the training process, random background music from MUSDB is overlaid onto clean vocal data to predict pitch and energy sequences. 
The HuBERT structure of the melody extractor is based on a pre-trained model~\cite{van2022comparison} using LibriSpeech-960~\cite{panayotov2015librispeech} corpus and, after surpassing 5000 steps, the HuBERT structure is frozen and no longer update.

For comparison, we introduce baseline1, also known as \textbf{``Origin Pitch\&Energy"}. 
The model of the baseline1 method is the same as the proposed, except that it does not include a HuBERT-based melody extractor; instead, it directly inputs pitch and energy into the melody encoder.
Additionally, to assess the model's robustness to noise, we use the approach of known as \textbf{``Separated Origin Pitch\&Energy"}(Baseline2), where songs are separated and then converted.

The baseline2 method utilizes the same model as baseline1, with the distinction that the pitch and energy are extracted from audio processed by a state-of-the-art (SoTA) music source separation model~\cite{rouard2023hybrid}.

All the audio files are down-sampled to 16 kHz. 
Mel-spectrogram is generated using a short-time Fourier transform (STFT) with a 50 ms frame size, a 10 ms frame hop, and a Hanning window function. 
Baseline and proposed systems are assessed under identical settings unless explicitly stated otherwise. 
A demo page is accessible\footnote{https://wei-chan2022.github.io/RobustSVC/}.



\vspace{-1pt}
\subsection{Overall Evaluation}
Table~\ref{tab:overall} presents the SVC capabilities of different models under noisy and clean vocal conditions. 
15 source audios that come from different source speakers and contents are used. 
To simulate noisy audio, random background music from MUSDB is overlaid onto the clean vocal data. 
An overall evaluation from both objective and subjective perspectives is conducted. 
The objective indicator includes F0 RMSE, while the subjective indicators encompass mean opinion scores (MOS) tests to assess naturalness and similarity.
F0 RMSE is used to measure the naturalness of the converted audio~\cite{huang2023singing}. 
F0 sequences are normalized using min-max normalization before processing. 
We compute RMSE between the converted and source waveform without BGM since the source waveform is originally performed by real singers with fine-grained melody and high naturalness. 
A lower RMSE value typically signifies a higher level of naturalness and intonation in the converted waveform. 
In the noisy and clean vocal condition, RobustSVC demonstrates excellent performance in terms of F0 RMSE, outperforming baseline1 (Origin Pitch\&Energy) and even surpassing baseline2 (Separated Origin Pitch\&Energy).
In the clean vocal condition, baseline2 exhibits higher F0 RMSE compared to baseline1, implying lower naturalness. 
We speculate that this is due to the separated signal used directly by the baseline2 losing some voice information, which leads to the instability of SVC.

The standard 5-scale MOS tests results are also shown in Table~\ref{tab:overall}. 
RobustSVC get the highest scores and significantly improved similarity and naturalness compared to the other two models in both conditions.

Overall, our method performs exceptionally well in both subjective and objective experiments conducted in noisy and noise-free environments.

\begin{table}[t]
  \caption{Overall evaluation. MOS results are reported with 95\% confidence intervals. }
  \label{tab:overall}
  \centering
  \begin{tabular}{cccc}
    \toprule
    \textbf{Model}      &  \textbf{Indicator} &\textbf{Noisy Vocal}   & \textbf{Clean Vocal}               \\
    \midrule
        \multirow{3}{*}{\makecell{Origin\\Pitch\&Energy}} & F0 RMSE&0.270  & 0.123  \\
    &naturalness & 2.46  ± 0.17 &   3.22  ± 0.15  \\
    &similarity & 2.55  ± 0.19 &    3.27  ± 0.15      \\
    \midrule
    \multirow{3}{*}{\makecell{Separated\\Origin\\Pitch\&Energy}}    & F0 RMSE& 0.202& 0.135 \\
    &naturalness &3.50  ± 0.17 &   3.44  ± 0.16  \\
    &similarity & 3.44  ± 0.15&    3.48  ± 0.15      \\
    \midrule
    \multirow{3}{*}{RobustSVC}   & F0 RMSE&\textbf{0.193} &\textbf{0.109}  \\
    &naturalness & \textbf{3.70 ± 0.17} &  \textbf{ 3.93 ± 0.16}  \\
    &similarity &\textbf{3.62 ± 0.16} &     \textbf{3.88 ± 0.14}     \\
    \bottomrule
  \end{tabular}
\end{table}

\begin{table}[t]
\vspace{5pt}
  \caption{Experiment on melody feature input.}
  \label{tab:word_styles}
  \centering
  \begin{tabular}{lc}
    \toprule
    \textbf{Melody feature type}      & \textbf{F0 RMSE }                \\
    \midrule
    None                   & 0.253                                 \\
    HuBERT Raw              & 0.167                          \\
    proposed                   & 0.066                               \\
    Pitch\&Energy          & \textbf{0.059}                    \\
    \bottomrule
  \end{tabular}
  \vspace{-8pt}
\end{table}

\subsection{Melody Feature Evaluation}


To facilitate a comparative analysis of diverse melody features, we integrate several representations extracted from clean audio into the model framework illustrated in Figure~\ref{fig:baseline}. 
Specifically, pitch\&energy sequences, HuBERT Raw feature which is from the pre-trained model\footnote{https://github.com/bshall/hubert} published in~\cite{van2022comparison}, and our novel melody feature are incorporated. 
The evaluation of F0 RMSE served as a metric to gauge the naturalness of the generated audio output.
As delineated in Table~\ref{tab:word_styles}, the absence of any melody feature input results in a complete failure of melody modeling by the conversion model. 
Despite some rhythmic information preservation within the self-supervised representations, the efficacy of modeling song melody information using the HuBERT Raw feature is notably constrained. 
Conversely, Our proposed melody feature, while slightly compromised in sound quality due to timbre leakage~\cite{chen2022large}, contains richer melody information, performing comparably to pitch\&energy sequences input.

\subsection{Noise Inputs Evaluation}
\vspace{3pt}

To further assess the model's noise robustness, we evaluate the accuracy of melody modeling under varying signal-to-noise ratios (SNRs). 
F0 RMSE is computed to gauge the naturalness and pitch accuracy of the generated audio output. 
In Table~\ref{tab:noise}, with the increase in noise level, all models show a decline in performance. 
Baseline1 (Origin Pitch\&Energy) is sensitive to noise, with even slight increases in noise level leading to considerable performance deterioration.
However, RobustSVC only experiences a slight decrease in performance as the noise level increases, with this decline being smaller than that observed in the other two models.

At higher noise levels (SNR$\leq$10), the model directly utilizing pitch\&energy as melody feature is notably affected, resulting in substantial melody degradation. 
RobustSVC demonstrates superior noise robustness compared to baseline2 (Separated Origin Pitch\&Energy), underscoring the benefits of integrating robust feature extracted from a HuBERT-based melody extractor.
When SNR exceeds or is equal to 15, where background noise is significantly lower than the human voice, the modeling performance of the four models is relatively similar.
        
  
  

\begin{table}[t]
  \caption{F0 RMSE in different model noise inputs. P\&E means Pitch\&Energy.}
  \label{tab:noise}
  \centering
  \scalebox{0.97}{
    \begin{tabular}{lccc}
        \toprule
        \textbf{Noise Level}      & \textbf{\makecell{Origin P\&E} }  & \textbf{\makecell{Separated\\Origin P\&E}} & \textbf{RobustSVC}               \\
        \midrule
        
        SNR=0                   & 0.309 &0.173 &\textbf{0.158}                                  \\
        SNR=5              & 0.254&0.159   &\textbf{0.148}                          \\
        SNR=10                    & 0.227 &0.147   &\textbf{0.129}                              \\
        SNR=15          & 0.176 &0.127   &\textbf{0.123}                   \\
        \bottomrule
      \end{tabular}
  }
\end{table}

\begin{table}[t]
\vspace{5pt}
  \caption{Ablation study.}
  \label{tab:ablation}
  \centering
  \begin{tabular}{lcc}
    \toprule
    \textbf{Model}      & \textbf{COS-SIM }  & \textbf{ F0 RMSE }            \\
    \midrule
    RobustSVC                  & \textbf{0.823} &\textbf{0.1017}                                 \\
    w/o $L_{rf}$            & 0.743 &0.1667                         \\
    w/o  CIN                  & 0.651 &0.3020                              \\
    w/o  $L_{sim}$         & 0.583 &0.3481                    \\
    Pitch\&Energy         & 0.663 &0.1022                             \\
    \bottomrule
  \end{tabular}
\end{table}

\subsection{Ablation Study}
\vspace{3pt}
To understand the contribution of each component to the model, F0 RMSE is computed between the converted and source waveform, and the speaker embeddings of the converted and real target audio are extracted using a pre-trained speaker verification model~\cite{wang2023wespeaker} for COS-SIM, where a higher value signifies greater similarity. 
Results in Table~\ref{tab:ablation} show that all components have a significant impact on our proposed model.
Model without $L_{sim}$ and CIN module leads to significant performance degradation in similarity and quality because both components could eliminate timbre information of the source speaker leaked from the HuBERT-based melody extractor.
In addition, the removal of the Real/Fake discriminator led to a degradation in speech quality.

\vspace{-4pt}
\section{Conclusions}

In this work, we introduce RobustSVC, a novel SVC architecture with an adversarial training strategy for any-to-one conversion. 
Our approach significantly improves the model's noise robustness compared to previous recognition-synthesis based approaches, enabling the conversion of audio with noise such as BGM into clean vocals performed by the target singer.
Experimental results demonstrate that the proposed RobustSVC outperforms the baseline methods in terms of both naturalness and similarity, covering scenarios with both noise and no noise. 
The ablation study is carried out to confirm the effectiveness of components in the RobustSVC model.
Looking ahead, our future focus will be on any-to-many SVC tasks and exploring the utilization of additional external unannotated and internal corpora to minimize naturalness loss and enhance the model's generalization.

\newpage
\section{Acknowledgements}

This work is supported by National Natural Science Foundation of China (62076144) and Shenzhen Science and Technology Program (WDZC20220816140515001, JCYJ20220818101014030).

\bibliographystyle{IEEEtran}

\bibliography{mybib}


\end{document}